\documentclass[conference]{IEEEtran}
\IEEEoverridecommandlockouts
\usepackage{cite}
\usepackage[linesnumbered,ruled,vlined]{algorithm2e}
\usepackage{amsmath,amssymb,amsfonts}
\usepackage{algorithmic}
\usepackage{graphicx}
\usepackage{textcomp}
\usepackage{xcolor}
\usepackage{tikz}
\usepackage{array}
\usepackage{threeparttable}
\usepackage{booktabs}
\usepackage{multirow}
\usepackage{makecell}
\usepackage{colortbl} % 用于设置背景颜色
\usepackage{subcaption} % 用于创建子图
\def\BibTeX{{\rm B\kern-.05em{\sc i\kern-.025em b}\kern-.08em
    T\kern-.1667em\lower.7ex\hbox{E}\kern-.125emX}}

\begin{document}

\title{DS-CIM: \underline{D}igital \underline{S}tochastic \underline{C}omputing-\underline{I}n-\underline{M}emory Featuring Accurate OR-Accumulation via Sample Region Remapping for Edge AI Models}

\vspace{-5mm}
\author{
    \IEEEauthorblockN{Kunming Shao\textsuperscript{1,*,†}, Liang Zhao\textsuperscript{2,3,*}, Jiangnan Yu\textsuperscript{1,*}, Zhipeng Liao\textsuperscript{4},\\ Xiaomeng Wang\textsuperscript{1}, Yi Zou\textsuperscript{2,3}, Tim Kwang-Ting Cheng\textsuperscript{1}, Chi-Ying Tsui\textsuperscript{1}}
    
    \IEEEauthorblockA{\textsuperscript{1}The Hong Kong University of Science and Technology,
    \textsuperscript{2}South China University of Technology,\\
    \textsuperscript{3}MoE Engineering Research Center of Design and Technology Co-Optimization of IC,
    \textsuperscript{4}Westlake University\\
    *Equally Contributed Authors. †Email: kshaoaa@connect.ust.hk}
    
    \thanks{This research was supported by ACCESS - AI Chip Center for Emerging Smart Systems, sponsored by InnoHK funding, Hong Kong SAR.}

    \vspace{-7mm}
}

\maketitle

\begin{abstract}
Stochastic computing (SC) offers hardware simplicity but suffers from low throughput, while high-throughput Digital Computing-in-Memory (DCIM) is bottlenecked by costly adder logic for matrix-vector multiplication (MVM). To address this trade-off, this paper introduces a digital stochastic CIM (DS-CIM) architecture that achieves both high accuracy and efficiency. We implement signed multiply-accumulation (MAC) in a compact, unsigned OR-based circuit by modifying the data representation. Throughput is enhanced by replicating this low-cost circuit 64 times with only a 1$\times$ area increase. Our core strategy, a shared Pseudo Random Number Generator (PRNG) with 2D partitioning, enables single-cycle mutually exclusive activation to eliminate OR-gate collisions. We also resolve the 1s saturation issue via stochastic process analysis and data remapping, significantly improving accuracy and resilience to input sparsity. Our high-accuracy DS-CIM1 variant achieves 94.45\% accuracy for INT8 ResNet18 on CIFAR-10 with a root-mean-squared error (RMSE) of just 0.74\%. Meanwhile, our high-efficiency DS-CIM2 variant attains an energy efficiency of 3566.1 TOPS/W and an area efficiency of 363.7 TOPS/mm$^{2}$, while maintaining a low RMSE of 3.81\%. The DS-CIM capability with larger models is further demonstrated through experiments with INT8 ResNet50 on ImageNet and the FP8 LLaMA-7B model.
\end{abstract}

% \begin{IEEEkeywords}
% Stochastic Computing, Digital Computing-in-Memory, Approximate Computing, AI Accelerators.
% \end{IEEEkeywords}

\section{Introduction}

Deep neural networks (DNNs) and Transformer models are foundational to modern AI, but their computational demands present significant challenges for energy- and area-constrained edge devices \cite{RestNet,girshick2014rich,krizhevsky2012imagenet,ren2016faster,vaswani2017attention, touvron2023llama, guo2025deepseek}. Efficient matrix vector multiplication (MVM) is critical for their edge deployment. Approximate computing has been explored to improve efficiency, but the high accuracy requirements of Transformers limit the applicability of many methods.

\begin{figure}[tp]
\centerline{\includegraphics[width=\columnwidth]{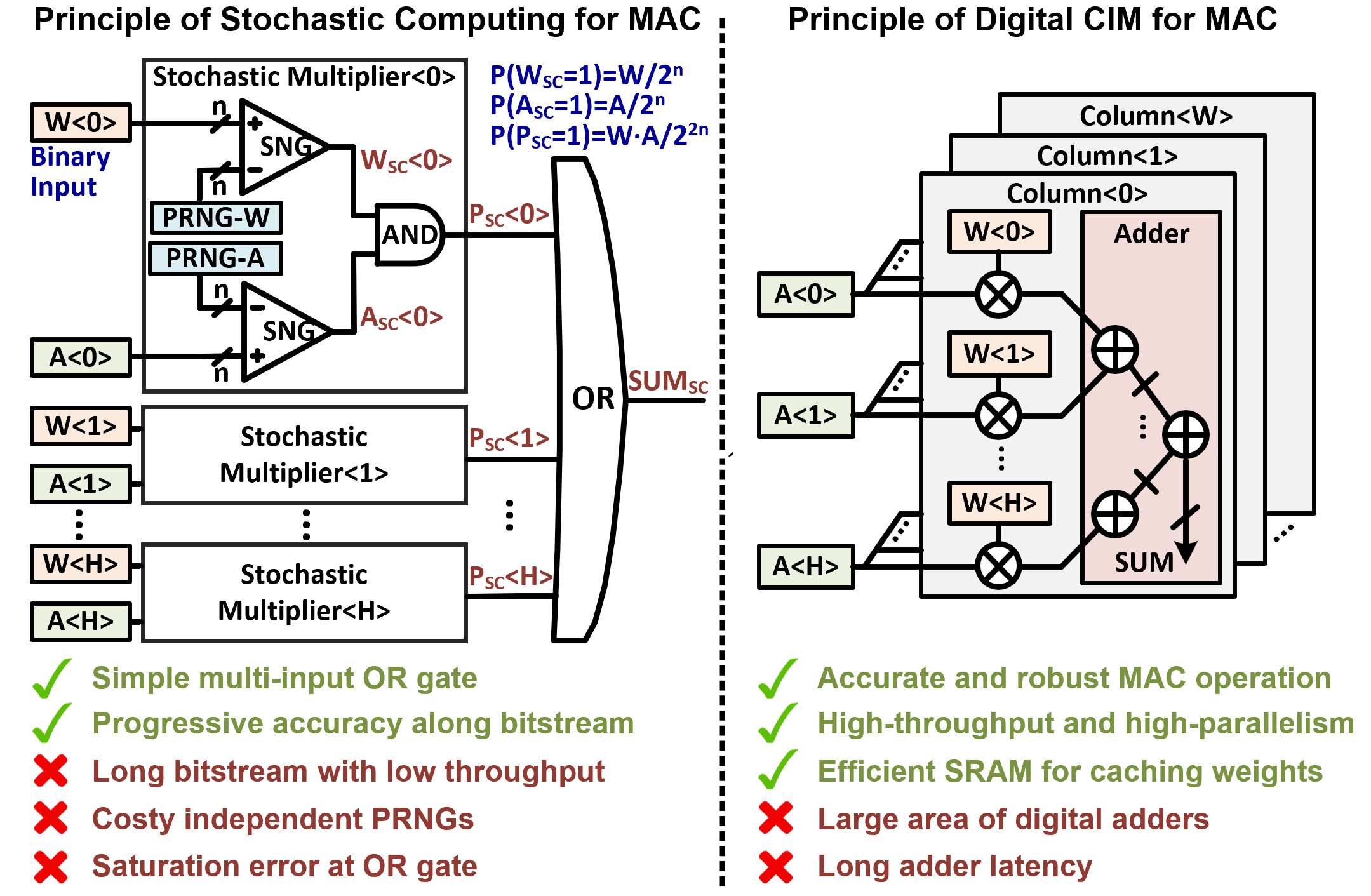}}
\vspace{-2mm}
\caption{\small{Principles of Stochastic Computing and Digital CIM.}}
\vspace{-5mm}
\label{fig1}
\end{figure}

As illustrated in Figure \ref{fig1}, OR-Gate-based stochastic computing (SC) is one of these approximate methods that offers low-cost and fault-tolerant computation. However, it is hampered by low throughput from long bitstream latencies and significant 1s saturation errors \cite{schober2021high} during low-sparsity accumulation, restricting its use in modern AI applications \cite{arXiv2024StochaPSUMIMC,TNano2024StochaMAC,frasser2022SC,agrawal2020revisiting,alaghi2014fast,zhang2024pacim}. Digital compute-in-memory (DCIM) is another method to enhance edge MVM by performing parallel computations directly in the SRAM array and reusing the weight, but its efficiency is often bottlenecked by the large and power-hungry digital adder tree used for accurate accumulation \cite{ISSCC2021dcim,ISSCC2022dcimDVFS,ISSCC2024dcimINT12}.

Analog CIM (ACIM) \cite{yoshioka202434, ueyoshi2022diana} and approximate DCIM \cite{ISCAS2024FPcim, lin2023dimca, he20237} sacrifice precision by utilizing low-precision ADCs or approximate adders for improved efficiency. ACIM employs efficient but PVT-sensitive ADCs for accumulation, while approximate DCIM uses simplified adders \cite{lin2023dimca} that introduce logical errors. However, their accuracy cannot meet the requirement of transformer models.

SC can achieve much lower errors by increasing the bitstream length. Integrating SC with CIM is a promising alternative, yet existing works suffer from inefficiencies with large weight bit-widths, limited parallelism, and persistent saturation errors, particularly with signed operations \cite{GLSVLSI2023stochastic,TCASII2022Stocha,agwa2023digital,yang202265nm,yang2024278}.

To address these challenges, we propose our DS-CIM architecture. \textbf{First}, we introduce a novel method that shifts signed numbers to an unsigned representation, allowing for efficient, unipolar stochastic computing that supports both signed activations and weights. \textbf{Second}, we replicate a simple unipolar OR-MAC circuit to enhance parallelism by 64$\times$, compensating for the throughput degradation from long bitstreams with only a 1$\times$ area increase. \textbf{Third}, we use a shared PRNG and a data remapping technique to ensure each random sampling point activates at most one row, completely eliminating 1s saturation errors. \textbf{Finally}, we optimize the PRNG configuration to minimize overall RMSE.

Post-layout simulations in a 40nm process validate our approach. Our precise DS-CIM1 achieves 669.7 TOPS/W and 117.1 TOPS/mm$^{2}$ with only 0.74\% RMSE. The efficient DS-CIM2 reaches 3566.1 TOPS/W and 363.7 TOPS/mm$^{2}$ with 3.81\% RMSE. DS-CIM demonstrates high efficiency, low error, and robust accuracy across all product sparsity levels (0-100\%).

% The remainder of this paper is structured as follows: Section II provides preliminaries. Section III presents the hardware architecture. Section IV details error analysis and optimization. Section V showcases experimental results. Finally, Section VI concludes the paper.

\section{Preliminaries}
%\textcolor{red}{Add more figures for the preliminaries, and more references}

\subsection{Stochastic Computing}

Stochastic computing generates unary bitstreams by comparing binary numbers with random values, ensuring each position's probability of a '1' matches the binary value. This conversion simplifies high-bit-width binary values, reducing logical computation costs and enhancing hardware efficiency\cite{arXiv2024StochaPSUMIMC,frasser2022SC,agrawal2020revisiting,alaghi2014fast,zhang2024pacim}.

Classic SC applications, such as multiplication and scaling, leverage the probabilistic nature of unary bitstreams for simpler, scalable hardware implementations. However, generating long unary bitstreams can increase latency, requiring careful management of the trade-off between efficiency and computation time based on application needs. There are also challenges that when applying SC on MAC operations, there will be distortion error due to the accumulation \cite{TNano2024StochaMAC, schober2021high}.

\subsection{Digital Computing-in-Memory}

AI accelerators face the memory wall problem, where memory access speed limits overall performance. Digital Compute-In-Memory (CIM) addresses this by integrating processing within memory units, reducing data transfer needs and enhancing efficiency\cite{ISSCC2021dcim,ISSCC2022dcimDVFS,ISSCC2024dcimINT12}.
However, challenges remain, such as low computing density and throughput, and the significant overhead of adder trees. Overcoming these issues is crucial for maximizing the potential of Digital CIM in AI accelerators.

\subsection{Approximate Computing-in-Memory}

Some works have applied approximate computing techniques to CIMs, such as using analog signals for accumulation \cite{yoshioka202434}, designing approximate adders \cite{lin2023dimca}, and introducing stochastic computing \cite{yang2024278}. However, these approaches face accuracy issues. Although high Root Mean Square Error can show good results in DNNs or event cameras, with the development of transformers, previous approximate computing methods are no longer applicable.

\section{DS-CIM Hardware Architecture}

\begin{figure*}[!t]
\centerline{\includegraphics[width=0.95\textwidth]{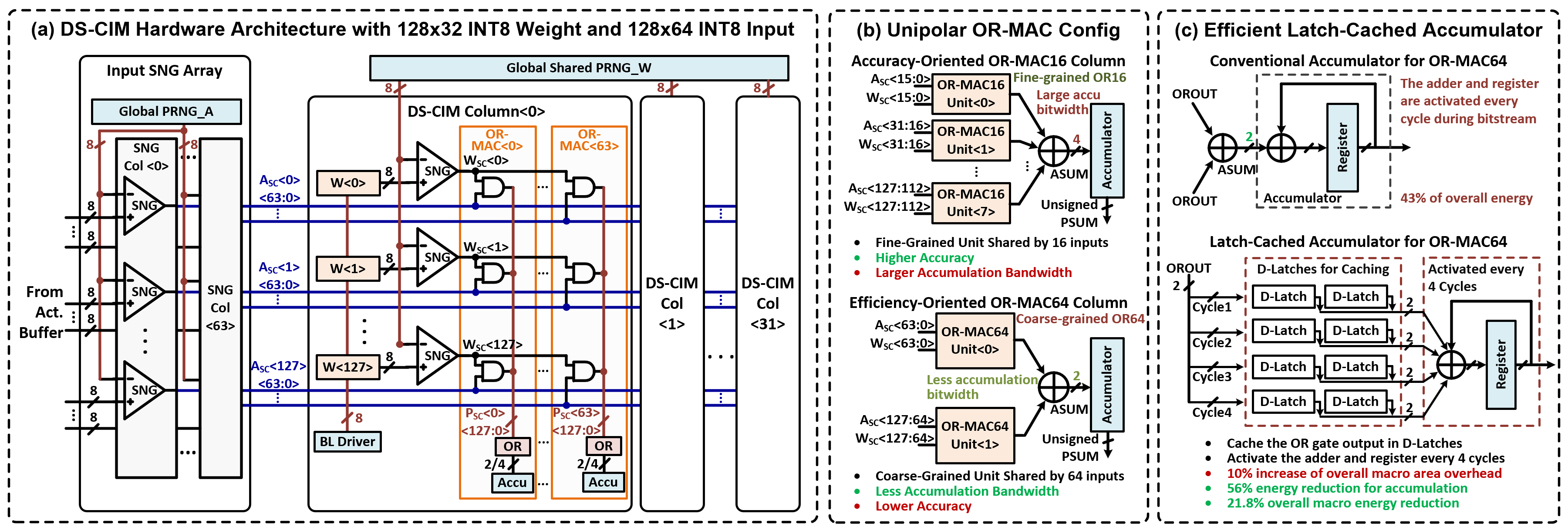}}
\vspace{-2mm}
\caption{\small{(a) Overall architecture of proposed DS-CIM. (b) Demonstration of the proposed unipolar OR-MAC configurations with the accuracy-oriented OR-MAC16 configuration for DS-CIM1 and the efficiency-oriented OR-MAC64 configuration for DS-CIM2. (c) Effcient latch-cached accumulator for reducing the high energy consumption of the conventional accumulators.}}
\vspace{-5mm}
\label{fig2}
\end{figure*}

\subsection{Overall Architecture}

% As shown in Figure \ref{fig2}, the DS-CIM architecture integrates the PRNGs and Stochastic Number Generators (SNGs) of SC into a DCIM framework, replacing costly multi-bit adders with efficient single-bit OR-based Multiply-Accumulate (OR-MAC) circuits.

As shown in Fig. \ref{fig2}(a), DS-CIM integrates SC's PRNGs and Stochastic Number Generators (SNGs) into a DCIM framework, replacing costly multi-bit adders with efficient single-bit  OR-based Multiply-Accumulate (OR-MAC) circuits. The macro has a 128x32 structure with 128 8-bit SRAMs, 128 SNGs, and 64 OR-MACs per column. An input SNG array converts binary activations to stochastic bitstreams. To support data remapping (Sec. IV), the macro uses only two independent PRNGs: PRNGA for activation bitstreams and PRNGW for weight bitstreams. These PRNGs generate random numbers for the SNGs to convert binary data into stochastic bitstreams, \(A_{SC}\) and \(W_{SC}\). \(A_{SC}\) is broadcast across columns, while \(W_{SC}\) is shared among the 64 OR-MACs within a column.

After multiplication, product bitstreams (\(P_{SC}\)) are accumulated via an OR gate, and the partial sum is added cycle-by-cycle in an accumulator. For balancing accuracy and efficiency, we designed two versions: the precise DS-CIM1 uses eight OR16-based MACs (OR-MAC16) per column, while the efficient DS-CIM2 uses two OR64-based MACs (OR-MAC64). Both perform accurate MAC operations but differ in their accumulation and data remapping granularities.

% The independent PRNGA and PRNGW generate two random numbers each cycle, forming a random sampling point. SNGs convert binary data into stochastic output bitstreams by comparing the random numbers from the PRNGs with the binary data. In this stochastic number generation process, binary activations and weights are transformed into bitstreams \(A_{SC}\) and \(W_{SC}\). The \(A_{SC}\) bitstream is broadcast across the DS-CIM columns, while the \(W_{SC}\) bitstream is shared among the 64 OR-MAC circuits in one column. After multiplication, the product bitstreams (\(P_{SC}\)) from different rows are accumulated through an OR gate, with the partial sum added cycle by cycle in an accumulator.

% After carefully considering the trade-off between accuracy and efficiency, we design two versions of DS-CIM with distinct accuracies and efficiencies: the precise DS-CIM1, which utilizes eight OR16 gate-based MAC circuits (OR-MAC16) in one column, and the efficient DS-CIM2, which employs two OR64 gate-based MAC circuits (OR-MAC64) in one column. Both OR-MAC16 and OR-MAC64 are capable of performing accurate MAC operations, although their accumulation and data remapping granularities differ.

\subsection{Unipolar OR-MAC for Signed MAC}

\begin{figure}[tp]
\centerline{\includegraphics[width=\columnwidth]{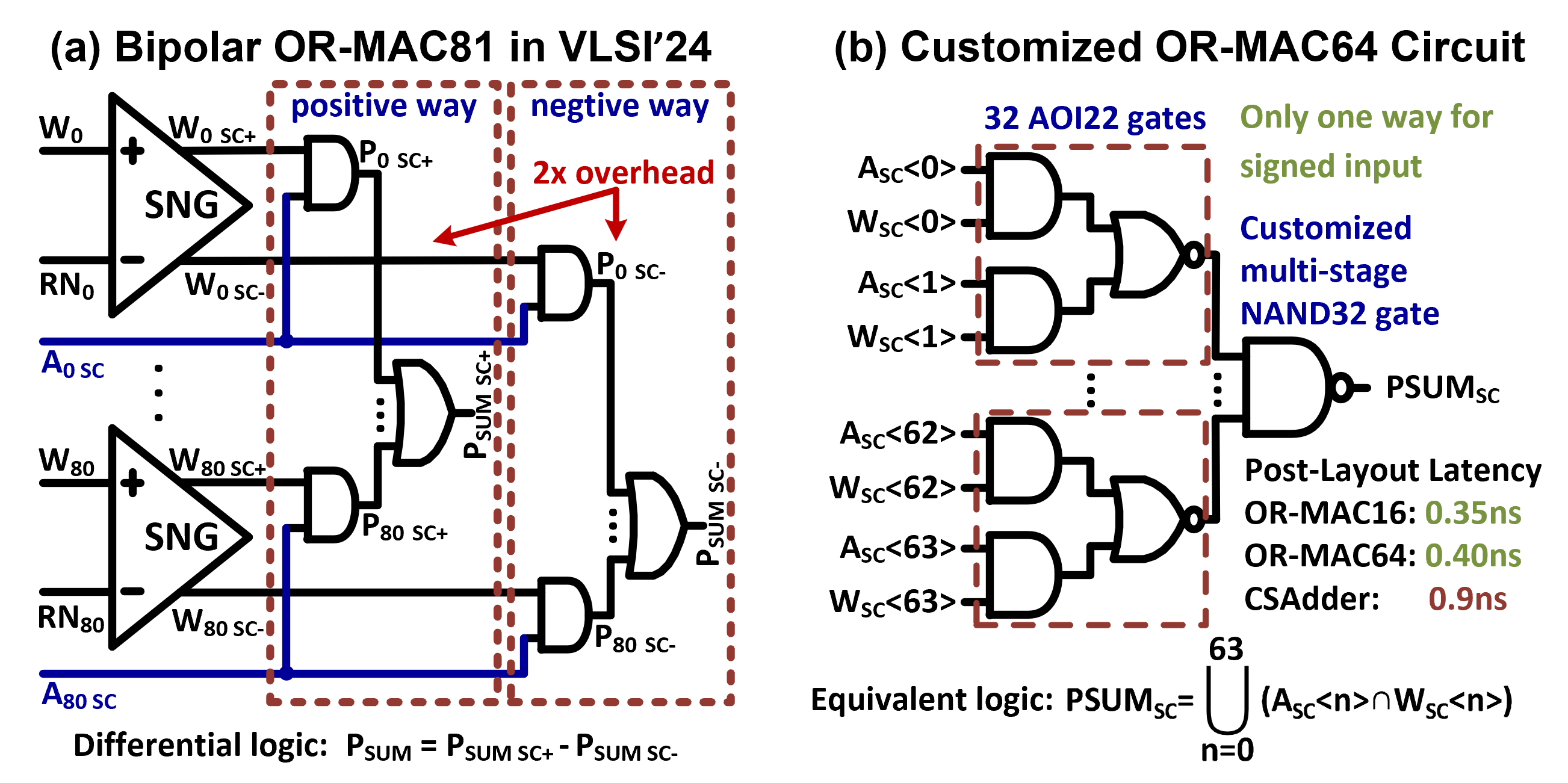}}
\vspace{-3mm}
\caption{\small{(a) Bipolar OR-MAC81 for event camera in \cite{yang2024278}. (b) Proposed unipolar OR-MAC64 circuit with only 0.4ns latency in 40nm process.}}
\vspace{-3mm}
\label{fig3}

\end{figure}

SC employs a probabilistic process to perform unsigned operations. As depicted in Figure \ref{fig3}(a), for executing signed MAC in stochastic CIM (S-CIM), \cite{yang2024278} proposes a sign-aware bipolar MAC circuit for signed weights but unsigned activations. This differential circuit separates the positive weight bitstream \(W_{SC+}\) from the negative weight bitstream \(W_{SC-}\) and performs MAC operations in two ways, with the final result being the difference between the two operations. However, this bipolar MAC scheme significantly increases overhead and limits parallelism and efficiency.

In contrast, DS-CIM introduces a unipolar MAC scheme for both signed activations and weights by converting signed numbers into unsigned ones and converting the output to signed numbers after the unsigned MAC. As shown in Eq. \ref{eq1}, the partial sum is logically derived from the MAC operation of two columns of 8-bit signed activations \(x[i]\) and weights \(w[i]\) represented in 2's complement. To convert a signed number to an unsigned one, we invert its sign bit, which is equivalent to adding 128. For example, we transform \(1000\_0011\) (2's complement, -125) to \(0000\_0011\) (unsigned, 3) by inverting the sign bit, and \(0111\_1111\) (2's complement, 127) to \(1111\_1111\) (unsigned, 255). Eq. \ref{eq2} and \ref{eq3} illustrate our method for decomposing signed multiplication into three terms, while Eq. \ref{eq4} demonstrates how we partition the high-cost signed MAC into efficient unsigned MAC, addition, and subtraction.

\begin{equation}
\centering
\begin{aligned}
psum = \sum_{i=0}^{H-1} x[i]w[i]
\end{aligned}
\vspace{-0.2cm}
\label{eq1}
\end{equation}
\begin{equation}
\begin{aligned}
x'[i] = x[i] + 2^7 \hspace{0.2cm} and \hspace{0.2cm}
w'[i] = w[i] + 2^7
\label{eq2} 
\end{aligned}
\end{equation}
\begin{equation}
\centering
\begin{aligned}
x[i]w[i] = x'[i]w'[i] - 2^7 x[i] - 2^7 w'[i]
\label{eq3} 
\end{aligned}
\end{equation}
\begin{equation}
\begin{aligned}
psum^{\tikz[baseline=-.7ex]{\node[draw, red, circle, inner sep=0.3pt] {\textcolor{red}{\scalebox{0.5}{a}}}}}
 = \sum_{i=0}^{H-1}x'[i]w'[i]^{\tikz[baseline=-.7ex]{\node[draw, red, circle, inner sep=0.3pt] {\textcolor{red}{\scalebox{0.5}{b}}}}}
 - 2^7\sum_{i=0}^{H-1} x[i]^{\tikz[baseline=-.7ex]{\node[draw, red, circle, inner sep=0.3pt] {\textcolor{red}{\scalebox{0.5}{c}}}}}
 - 2^7\sum_{i=0}^{H-1} w'[i]^{\tikz[baseline=-.7ex]{\node[draw, red, circle, inner sep=0.3pt] {\textcolor{red}{\scalebox{0.5}{d}}}}}
\label{eq4}
\end{aligned}
\end{equation}

% \begin{equation}
% \begin{aligned}
% \scalebox{0.9}{\boxed{\raisebox{15pt}{\tikz[baseline]{\node[draw, red, circle, inner sep=0.5pt] {\textcolor{red}{\scalebox{1}{a}}}}} \sum_{i=0}^{H-1} X[i]W[i]}}
%  =& \scalebox{0.9}{\boxed{\raisebox{15pt}{\tikz[baseline]{\node[draw, red, circle, inner sep=0.5pt] {\textcolor{red}{\scalebox{0.8}{b}}}}} \sum_{i=0}^{H-1}X'[i]W'[i]}} - \scalebox{0.9}{\boxed{\raisebox{15pt}{\tikz[baseline]{\node[draw, red, circle, inner sep=0.5pt] {\textcolor{red}{\scalebox{1}{c}}}}} 128\sum_{i=0}^{H-1} X[i]}} \\
%  &- \scalebox{0.9}{\boxed{\raisebox{15pt}{\tikz[baseline]{\node[draw, red, circle, inner sep=0.5pt] {\textcolor{red}{\scalebox{0.8}{d}}}}} 128 \sum_{i=0}^{H-1} W'[i]}}
% \label{eq4}
% \end{aligned}
% \end{equation}
\vspace{-0.2cm}

In our DS-CIM hardware, term \(b\) can be efficiently calculated using our unsigned unipolar OR-MAC, term \(c\) is summed through SIMDs during runtime and term \(d\) is computed offline and stored as a look-up table. The overhead for term \(c\) is spatially amortized across multiple weight columns. Compared to the substantial MAC overhead, the subtraction and amortized addition overheads are negligible.

%\textcolor{red}{In this work, we customized the OR-MAC circuit, and we should show the circuit schematic to them in the experiment session, and latency analysis.}

Figure \ref{fig2}(b) illustrates our two unipolar OR-MAC circuit configurations. OR-MAC16 is accuracy-oriented, featuring finer OR granularity with a larger addition bitwidth, whereas OR-MAC64 is efficiency-oriented, characterized by coarser OR granularity and a smaller addition bitwidth. Figure \ref{fig3}(b) presents the schematic of our customized OR-MAC64 circuit, which is smaller and faster than carry-save adders \cite{shao2024syndcim}.

\subsection{Compute/Memory Ratio and the Area Efficiency}

%\textcolor{red}{Here we should clarify that why the latency could be reduced by 64x times, make the figure more clear}

To achieve \textit{n}-bit precision, SC requires a bitstream length of \(2^{n}\). However, for an 8-bit MAC operation, the delay of a 256-point SC is 32 times longer than that of an 8-bit serial DCIM, which only requires 8 cycles. Furthermore, in S-CIM that has one set of memory with only one set of MAC circuit, the overhead of the OR-MAC circuit is significantly lower than that of SRAMs and SNGs, leading to unbalanced computation and area efficiency.

\begin{figure}[!tp]
\centerline{\includegraphics[width=0.9\columnwidth]{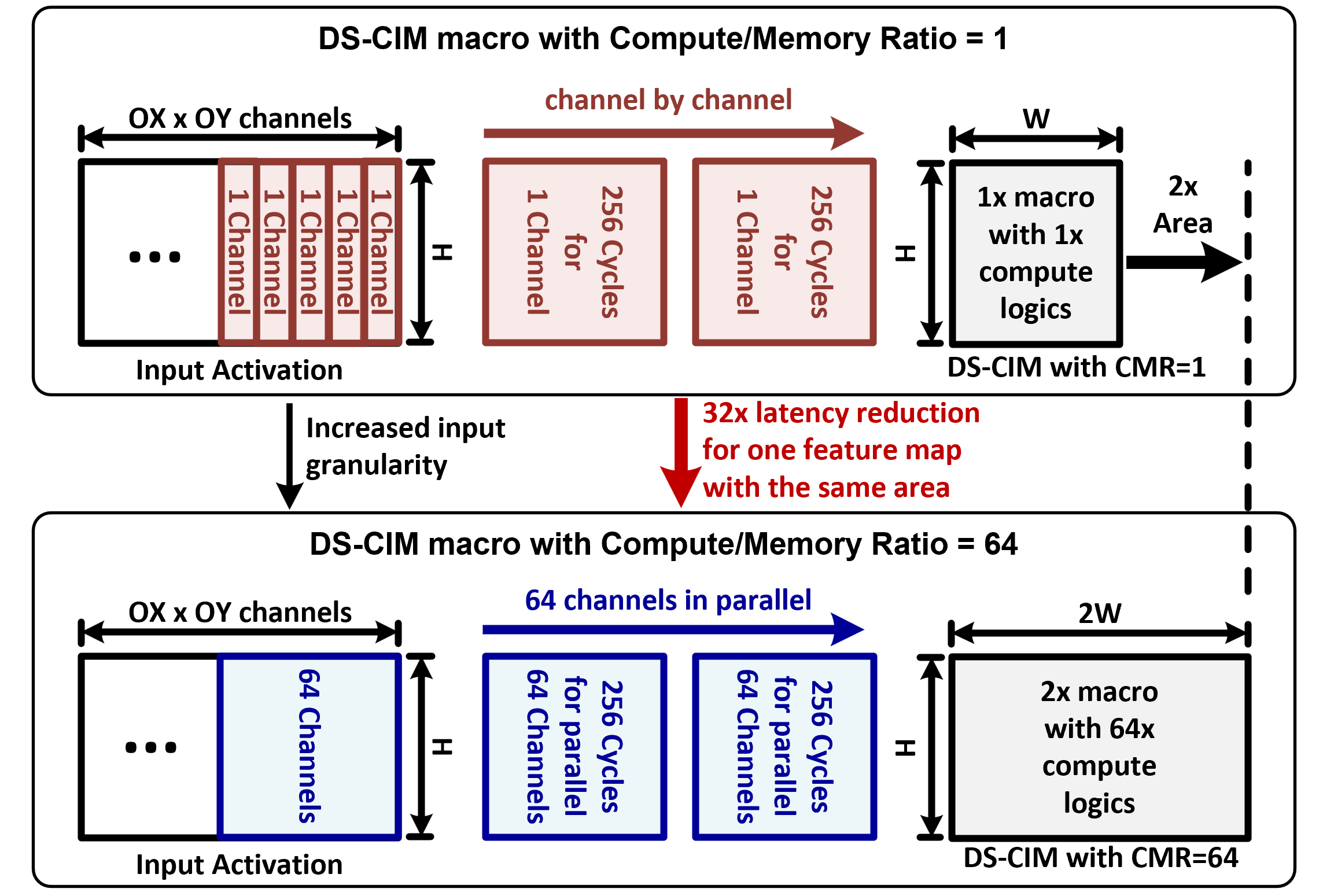}}
\vspace{-2mm}
\caption{\small{An illustration of achieving 32x latency reduction and 32x compute density increment by increasing CMR from 1 to 64.}}
\vspace{-3mm}
\label{fig4}
\end{figure}

To address this issue, we replicate the OR-MAC circuit within the same DS-CIM column \cite{yang2024278}. By integrating 64 ultra-efficient OR-MAC circuits in a column, denoted as compute/memory ratio (CMR) = 64, the parallelism of DS-CIM is enhanced by 64$\times$, boosting the throughput and area efficiency. Our post-layout data shows that DS-CIM with CMR=64 incurs only a 1$\times$ area increase compared to CMR=1 designs. As illustrated in Figure \ref{fig4}, By 64$\times$ throughput with only 2$\times$ area, DS-CIM achieves a 32$\times$ latency reduction for a specific CONV layer. Moreover, DS-CIM features progressive precision, enabling a 4$\times$ extra improvement with a bitstream length=64. By replicating low-cost unipolar OR-MAC circuits, we have mitigated the throughput limitations inherent in SC, providing higher throughput.

\begin{figure}[tp]
\centerline{\includegraphics[width=\columnwidth]{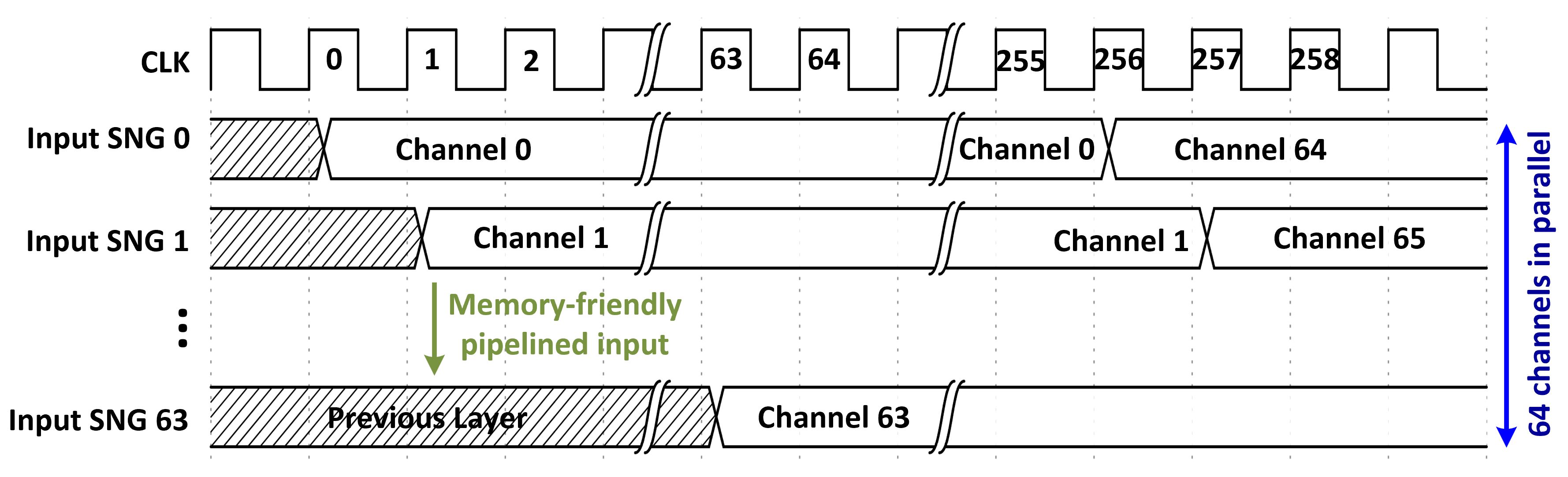}}
\vspace{-3mm}
\caption{\small{Memory-friendly pipelined layer channel input timing diagram of the DS-CIM macro with CMR=64 and bitstream length=256.}}
\vspace{-3mm}
\label{fig_new}
\end{figure}

As shown in Figure \ref{fig_new}, since the activations and weights are temporally stationary during SC process, the SRAM bandwidth could be efficiently utilized for different input channels.

\begin{figure*}[!t]
\vspace{-3mm}
\centerline{\includegraphics[width=0.95\textwidth]{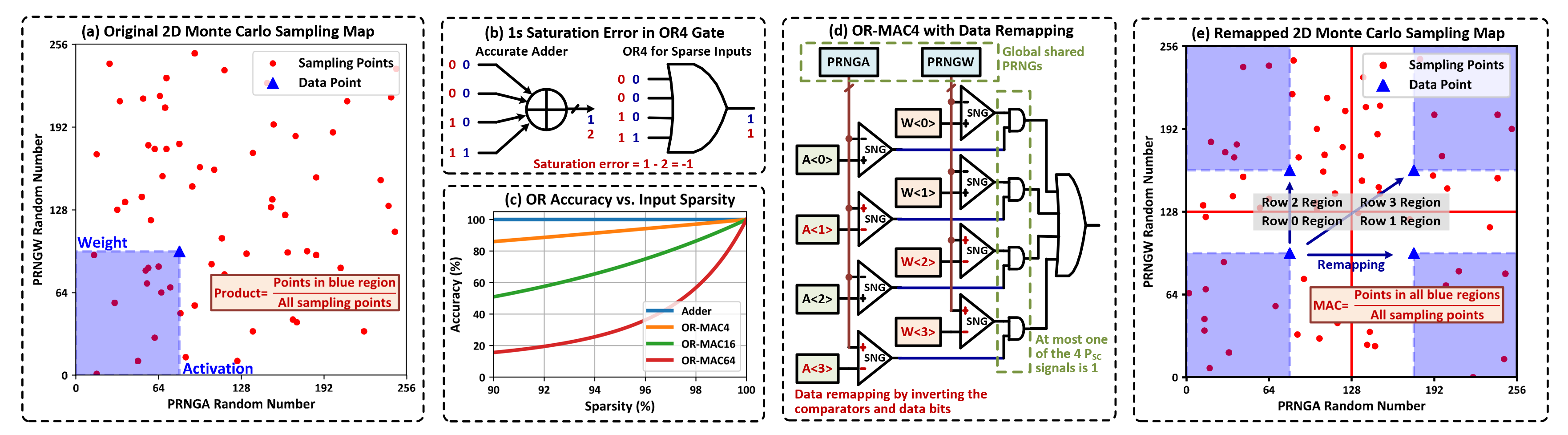}}
\vspace{-2mm}
\caption{\small{(a) Representing stochatisc multiplication as 2D Monte Carlo process. (b) 1s saturation error in the OR4 gate. (c) The relationship between conventional OR-MAC accuracy and the input product sparsity. (d) OR-MAC4 supporting data remapping with inverted data and comparators. (e) Data remapping in 2D Monte Carlo sampling map, where a sampling point appear at most in one of the four blue regions.}}
\vspace{-4mm}
\label{fig6}
\end{figure*}

%By replicating more low-cost unipolar OR-MAC circuits within a limited area budget, we have mitigated the throughput limitations inherent in stochastic computing, providing higher parallelism and throughput for AI acceleration.

\subsection{Latch-Cached Bitstream Accumulator}

As shown in Figure \ref{fig2}(c), DS-CIM requires accumulating a long bitstream cycle by cycle. In DS-CIM, due to the replication of low-cost OR-MAC units, the accumulator's energy consumption accounts for 43\% of the entire macro. For DS-CIM2, which has a smaller accumulation bitwidth, we propose a latch-cached accumulator. We put eight D-latches at the 2b output of the OR-MAC. The outputs are sequentially stored in latches, filling them every four cycles. On the fourth cycle, when the D-latches are full, an adder sums the four 2-bit cached numbers and accumulates the result with the previously stored value in registers. This approach activates the accumulator only once every four cycles, reducing accumulation energy by 56\% and overall macro power consumption by 21.8\%, with only 10\% increasement in area.

%\textcolor{red}{Here we demonstrate our data using sram/latch cached bitstream accumulator for reducing the huge accumulation overhead. }

%\textcolor{red}{And with one more figure for this subsection.}

%Section III Section III Section III Section III
\section{Error Analysis and Optimization}

% \begin{figure}[tp]
% \centerline{\includegraphics[width=\columnwidth]{figure/fig5.png}}
% %\vspace{-2mm}
% \caption{\small{(a) Representing stochatisc multiplication as 2D Monte Carlo process. (b) 1s saturation error in the OR4 gate. (c) The relationship between conventional OR-MAC accuracy and the input sparsity.}}
% \vspace{-3mm}
% \label{fig5}
% \end{figure}

In Stochastic CIM (S-CIM) architecture, due to the nature of random sampling and accumulation, large errors in each row can cancel each other out, resulting in a relatively smaller MAC error. However, inherent errors, such as 1s saturation error and sampling point discrepancies, introduce systematic errors that affect S-CIM. In this section, we analyze the error sources in our DS-CIM and propose corresponding solutions.

\subsection{Random Sampling and 1s Saturation Error}

Activation and weight multiplication can be viewed as a 2D Monte Carlo sampling process that estimates the product by calculating the area formed by the two values. As depicted in Figure \ref{fig6}(a), a red sampling point, \((PRNGA, PRNGW)\), appears randomly in the 2D sampling map cycle by cycle. When a sampling point falls within the blue rectangle defined by weight and activation, the conditions of the activation and weight SNG comparators are met, causing both \(A_{SC}\) and \(W_{SC}\) to equal 1, resulting in an output product of 1. Consequently, the number of 1s in the product bitstream \(P_{SC}\), which reflects the magnitude of the product of activation and weight, depends on how many random points fall within the rectangular area during the sampling process. Thus, the stochastic multiplication process is effectively a 2D Monte Carlo process \cite{alaghi2014fast}.

Next, we consider S-CIM, where multiple independent Monte Carlo sampling processes occur across different rows. Previous work \cite{yang2024278} relied on independent PRNGs for each row to generate independent \(P_{SC}\)s, which are then accumulated together by an OR gate. However, the OR gate is particularly sensitive to the sparsity of \(P_{SC}\). As shown in Figure \ref{fig6}(b), taking an OR4 gate as an example, if the input sparsity is high, most cycles will have all-0 inputs or just a single 1. Occasionally, there may be multiple inputs equal to 1. However, the OR gate can only output a single 1, leading to a \textbf{1s saturation error}. A detailed probabilistic analysis in Figure \ref{fig6}(c) reveals that OR accumulation is highly sensitive to product sparsity, with coarser-grained OR gates exhibiting even greater sensitivity to variations in sparsity, which restricts the applicability of OR gate-based S-CIM.
%（1）蒙特卡洛仿真原理
% 红色采样点在2D采样map里面随机出现，如果采样点出现在由Weight和Activation形成的一个蓝色矩形内时，activation和weight SNG comparators的判定条件同时被满足，Asc和Wsc将同时为1，并输出一个Psc=1. 因此，Psc里1的个数即Activation和weight的乘积大小取决于采样过程中有多少随机点落在矩形范围内，这是一个典型的2D monte carlo process [DATE'14].
%（2）以往的工作在一列中的每一行采用相互独立的PRNG，因此每一行Psc的产生也具有独立性，这导致了OR累加的一个1s saturation error
%（3）以一个OR4 gate为例，假设四个输入每个输入的Psc sparsity为99%，由概率分析可知四输入同时出现两个或以上1的概率为xxx%,而OR累加只能输出一个1，因此导致一个1s saturation error.我们进行深度的概率分析可知，OR累加对输入的sparsity非常敏感，且越粗粒度的OR gate对sparsity变化越敏感。这限制了SCIM的应用

\subsection{Data Remapping for 1s Saturation error}

To address the 1s saturation error of OR accumulation, we re-examine the random independence between rows and discover that sharing PRNG between rows allows us to map all data points to the same 2D sampling map. By shifting and symmetry, we ensure that each sampling point can activate at most one row to output 1.

Specifically, as shown in Figure \ref{fig6}(a), we replace the strategy of deploying independent PRNGs for each row in \cite{yang2024278} with a shared PRNGW for all weights and a shared PRNGA for all activations, enabling each row in the column to utilize the same sampling point. We illustrate this with the OR-MAC4 case. In Figure \ref{fig6}(e), we right-shift both activations and weights by 1 bit, causing the data distribution region enclosed by red lines to occupy only the \textit{lower-left quarter} of the 2D sampling map.

We then invert all bits of the Row 1's activation \(A[1]\), making it spatially equivalent to a symmetry of 127.5, e.g., from \(0000\_0010\) to \(1111\_1101\). Simultaneously, we alter the decision direction of the second row's activation SNG. This remaps the red data distribution region and blue Monte Carlo effective region of the second row from the \textit{lower-left corner} to the \textit{lower-right corner}. Similarly, we can position Row 2 in the \textit{upper-left corner} by inverting \(W[2]\) and its SNG, and Row 3 in the \textit{upper-right corner} by inverting \(A[3]\), \(W[3]\), and the corresponding two SNGs. For a shared sampling point, \textbf{it will lie in at most one of the four blue rectangles}, ensuring that at most one of the four inputs to the OR4 gate will be 1. By sacrificing data bit-width through right-shifting, we can eliminate the 1s saturation error.

Similarly, we can implement a 4$\times$4 division of the 2D sampling map by right-shifting by 2 bits and inverting the corresponding data bits, with every 16 rows sharing an OR16 gate, forming an OR-MAC16 design, denoted as precise \textbf{DS-CIM1}. Furthermore, by right-shifting by 3 bits and remapping the data, we can divide the 2D sampling map into 8$\times$8 segments, with every 64 rows sharing an OR64 gate, forming an OR-MAC64 design, denoted as efficient \textbf{DS-CIM2}.

%（1）为了解决OR gate accumulation的问题，实现高效低成本的accumulation，in DS-CIM, 我们重新审视了行与行之间的独立性，发现可以通过建立起行与行之间的相关性，来将所有行的data point映射到同一个2D sampling map中，然后通过缩放和对称，使得共享OR gate的每一行都独立的分布在2D sampling map的一个grid中，使得每个sampling point至多使OR gate的一个输入变为1.
%（2）具体来说，我们将每行独立的PRNG策略替换为一列activation或者weight共享一组PRNG，使得一列中的每一行都共享一个sampling point。Figure 6以OR4 gate为例，我们将activation和weight都右移一位，使得data的分布范围只占据2D sampling map中的1/4。接着我们invert第二行的activation的所有位，使其在空间上等效为关于128对称（如0000_0010 to 1111_1101），同时我们改变第二行activation SNG的判定方向，如果random number大于activation，则Asc=1.这样，我们第二行的数据分布范围和monte carlo判定范围就从左下角变换到了右下角。同理，我们可以将第三行变换到左上角，第四行变换到右上角。当random sampling point出现时，其至多只会出现在四个蓝色矩形其中之一，也就是说，OR4 gate的四个输入中至多只有一个1.
%（3）同理，我们可以通过右移两位和稍微更复杂些的data remapping来实现将2D sampling map进行16等分，每16行共享一个OR16 gate, 形成一个OR-MAC16设计，即DS-CIM1。进一步地，我们右移三位和data remapping，将2D sampling map进行64等分，每64行共享一个OR64 gate，形成一个OR-MAC64设计，即DS-CIM2.
%（3）As depicted in Figure 7, 我们实测发现数据右移不仅没有造成精度下降，而且由于右移和data remapping消除了OR accumulation的误差使得整体误差降低到一个很小的水平。我们通过低成本的OR累加实现了与精确加法器完全一致的结果。测试结果展示了我们的unipolar OR-MAC circuits是sparsity variation resilient的，即对任何sparsity的input和activation都展现出了均匀的误差，这使得我们的DS-CIM可以应用于具有任意数据分布的应用，提高了DS-CIM的兼容性

% \begin{figure}[!t]
% %\vspace{-3mm}
% \centerline{\includegraphics[width=\columnwidth]{figure/fig7.png}}
% \vspace{-2mm}
% \caption{\small{DS-CIM1 and DS-CIM2 error analysis for signed 8bit results with a bitstream length of 256, 128, and 64, respectively.}}
% \vspace{-2mm}
% \label{fig7}
% \end{figure}
\subsection{Measurement of Optimized RMSEs}
%DS-CIM1和DS-CIM2在不同bitstream长度下的误差分析。对于同一个DS-CIM，bitstream越长计算结果的误差越小。对于相同长度bitstream，DS-CIM1的计算误差小于DS-CIM2。

%As analyzed in \cite{alaghi2014fast}, a lower distribution discrepancy of random sampling points in the 2D sampling map results in better uniformity of the sampling points, which in turn leads to reduced errors in random calculations. Also, the types of PRNGs and the initial values of random sequences can significantly impact MAC errors, as sampling points may be overly dense in some regions while too sparse in others within the 2D sampling map.

The distribution uniformity of sampling points, which is affected by PRNG types and intitial seeds, is crucial to the SC accuracy\cite{alaghi2014fast}. To minimize the distribution discrepancy of sampling points, we collected mainstream 8-bit PRNGs and searched for optimal initial values for the two random number sequences of PRNGA and PRNGW. We identified optimal PRNG and initial value configurations for 64, 128, and 256 points that minimize the overall RMSE of OR-MAC16 and OR-MAC64 for different data distributions. These configurations ensure optimal RMSE for each application during runtime.

%---------------------Table I------------------------------------
\begin{table}[t]
    \centering
    \vspace{-2mm}
    \caption{DS-CIM Performance on CNN Models}
    \vspace{-2mm}
    \begin{threeparttable}
    \resizebox{\columnwidth}{!}{%
    \begin{tabular}{lcccccc}
    \toprule
     & \multicolumn{3}{c}{\textbf{DS-CIM1}} & \multicolumn{3}{c}{\textbf{DS-CIM2}} \\
    \cmidrule(lr){2-4} \cmidrule(lr){5-7}
    Bistream & 64 & 128 & 256 & 64 & 128 & 256 \\
    \midrule
    RMSE & 3.57\% & 2.03\% & 0.74\% & 3.81\% & 2.63\% & 0.84\% \\
    \midrule
    \makecell[l]{ResNet18@\\CIFAR10(vs 94.54\%)} & 90\% & 93.08\% & 94.45\% & 89.46\% & 92.46\% & 94.31\% \\
    \midrule
    \makecell[l]{ResNet50@\\ImageNet(vs 80.82\%)} & 79.45\% & 80.11\% & 80.67\% & 79.34\% & 79.9\% & 80.65\% \\
    \bottomrule
    \end{tabular}%
    }
    \end{threeparttable}
    \vspace{-5mm}
    \label{Table I}
\end{table}
%---------------------Table I------------------------------------

Table \ref{Table I} illustrates the RMSE performance of our different OR-MAC circuits and bitstream lengths. Precise DS-CIM1 with bitstream length=256 achieves the lowest RMSE of 0.74\% and Efficient DS-CIM2 with bitstream length=64 achieves the acceptable RMSE of 3.81\%, which is better than the 4.03\% and 6.76\% reported in \cite{lin2023dimca}. Due to greater precision loss from right-shifting, the RMSE of OR-MAC64 is slightly higher than that of OR-MAC16. Additionally, the RMSE significantly increases with shorter sequences due to greater discrepancies and insufficient sampling points.

The test results demonstrate that our unipolar OR-MAC circuits are resilient to variations in sparsity, exhibiting uniform errors for any magnitude distribution of the partial sum, making our DS-CIM compatible with MVM applications across flexible data distributions.
%（1）如[DATE‘14]分析，2D sampling map中的点的分布discrepancy越低，随机点的uniformity越好，随机计算的误差也会越低。我们在测试中发现，不同的PRNG类型和initial values会导致MAC计算的误差差异很大，这是由于采样点在某些区域过于密集，而在某些区域过于稀疏导致的。
%（2）如算法1所示，为了降低sampling points的distribution discrepancy，我们搜集了主流的8bit PRNG，并对PRNGA和PRNGW两个随机数序列的initial values进行搜索，分别获取64点、128点和256点使得OR-MAC16和OR-MAC64 overall RMSE最优的种子组合存放在DS-CIM macro上，保证每种精度配置下的最优RMSE。
%（3）Figure 8展示了我们在不同的OR-MAC电路和bitstream length下的RMSE情况。可以发现由于更大的右移精度损失，OR-MAC64比OR-MAC16的RMSE更大。而由于短序列更大的discrepancy和缺乏充足的采样点，RMSE随bitstream length的长度减少而明显升高。然而即使是RMSE最大的OR-MAC64 with a bitstream length of 64, 其RMSE仅有xxx\%,比[DIMCA]更优秀。

\section{Experiments and Validation}
%\textcolor{red}{The experiment setup should be clarified that why we use virtuoso, dc, and innovus.}

%\textcolor{red}{We should add local layout and global layout of the DS-CIM macro here.}

To demonstrate the advantages of DS-CIM in hardware efficiency, flexibility, and accuracy, we implemented the layout designs of DS-CIM1 and DS-CIM2 using a 40nm CMOS process. The customized SNGs and OR-MAC units were designed in Cadence Virtuoso and characterized into standard cells for synthesis in Cadence Liberate and Cadence Abstract, as the method introduced in \cite{shao2024syndcim}. After obtaining all the necessary digital standard cells, the digital gate-level netlist was synthesized using Synopsys Design Compiler. Placement and routing were executed in Cadence Innovus, and post-layout simulation was performed with Cadence Virtuoso. 

For software evaluation, we conducted comprehensive experiments across quantized DNN and transformer models to assess the impact of stochastic error on model accuracy. The DS-CIM error pattern was added to the MVM results of the models. We tested \texttt{ResNet18} model on \texttt{CIFAR-10} and a larger \texttt{ResNet50} on \texttt{ImageNet} for accuracy evaluation under INT8 quantization, respectively.
 
We also employed the open-source \texttt{LLM-FP4} quantization framework \cite{liu2023llm} to apply the FP8 quantization to the \texttt{LLaMA-7B} model, assessing DS-CIM's effectiveness on mainstream datasets. Additionally, following the method outlined in \cite{tu2022redcim}, FP8 activations and weights were aligned to INT8 with a granularity of 128 as inputs for DS-CIM.

%\textcolor{red}{Here, we should include the FP8 process, resnet50, and transformer models}
%(1) Why our error is so low:
%1. eliminating the OR-accumulation error
%2. Utilizes the inherent advantage of DCIM structure that the error from each row cancels each other.
%3. Fine-tuned PRNGs.

%(2) Why our efficiency is so high:
%1. High sparsity OR-MAC: in each OR gate, only one row is activated to output 1, the other rows are all 0.
%2. Only have two PRNGs, eliminating the high dynamic cost.

%\textcolor{red}{Hardware: power/area breakdown}
%\textcolor{red}{Hardware: Area comparison with digital adder:64x OR-MAC16 and OR-MAC64}
%\textcolor{red}{tradeoff: power/area efficiency vs. resnet18 accuracy}
%\textcolor{red}{}

\subsection{Hardware Evaluations}

\begin{figure}[tp]
%\vspace{-4mm}
\centerline{\includegraphics[width=\columnwidth]{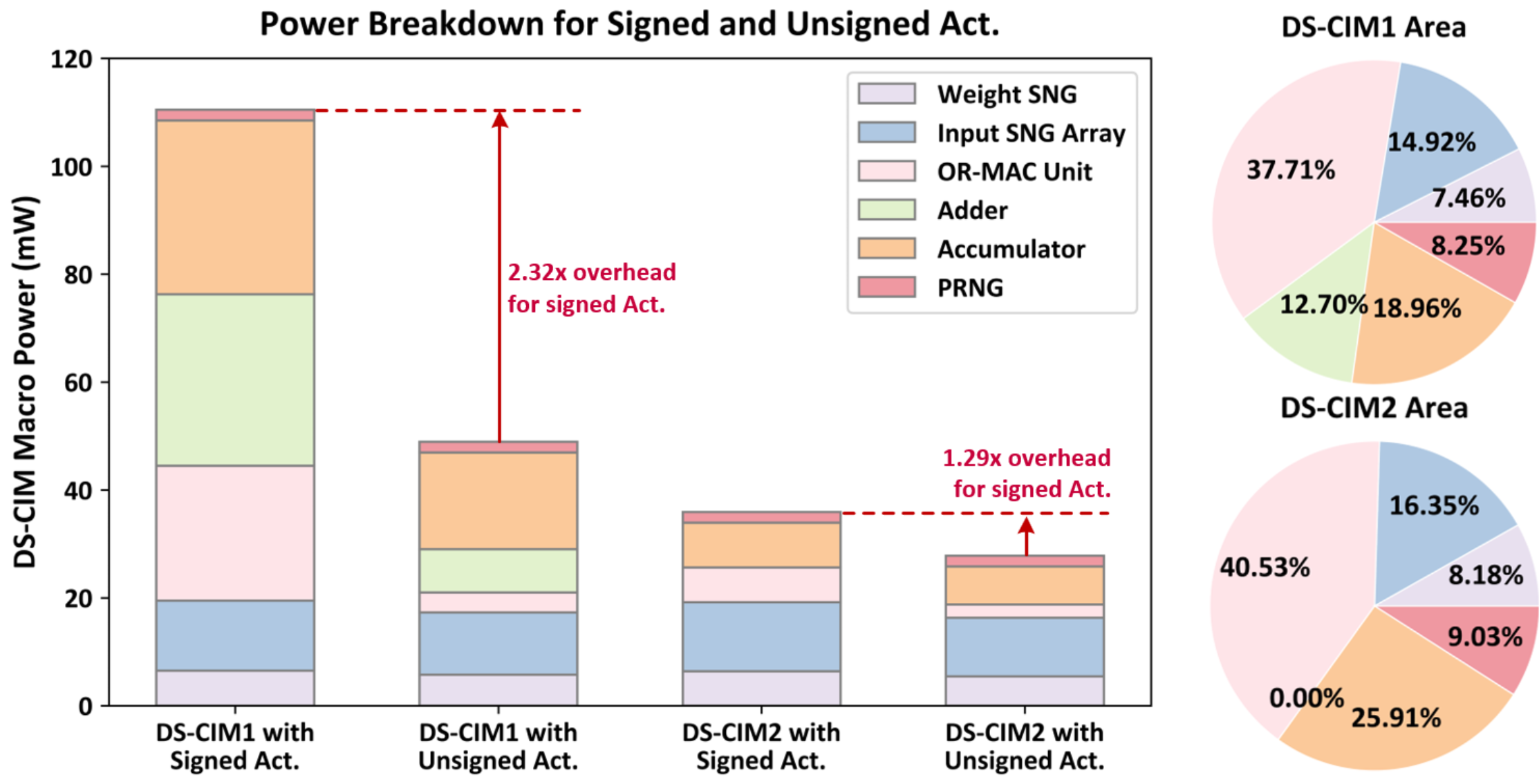}}
%\vspace{-2mm}
\caption{\small{Power and Area breakdown of DS-CIM1 and DS-CIM2 for signed and unsigned activations, where DS-CIM2 is equipped with the latch-cached accumulator.}}
%\vspace{-4mm}
\label{fig8}
\end{figure}

We first evaluate the power performance of the two designs, as well as their sensitivity to signed operations. As shown in Figure \ref{fig8}, the data remapping ensures that there is at most a single 1 in the input of each OR gate, achieving ultra-low power consumption. Due to the coarser OR-MAC granularity, simpler addition and lower accumulation bandwidth compared to DS-CIM1, DS-CIM2 is well compatible with latch-cached accumulators and consumes less energy in the MAC circuit than DS-CIM1. In signed operations, we transform the signed data to unsigned [0, 255], which reduces the sparsity of the stochastic bitstream, and increases power consumption because of more 1s to be accumulated.

As depicted in Figure \ref{fig8}, our evaluation indicates that DS-CIM1 is sensitive to signed operations, whereas DS-CIM2, with OR-MAC64 and simple addition logic, demonstrates resilience to signed operations. Additionally, we minimize the power consumption of PRNGs by reusing them for all weights and activations. However, SNGs and accumulators still contribute significantly to dynamic power consumption. For the area, we observe that a high portion of adders in DS-CIM1, which also confirms that increasing the granularity of the OR-MAC reduces the overhead of adder and accumulator logic, resulting in higher efficiency. Additionally, by duplicating the OR-MAC column 64 times, we have amortized the overhead of the weight SNGs, keeping its area and power proportions within a reasonable range.

%(0) 由于精巧的Data remapping设计，每个OR-MAC unit的输入中at most有一个1，这使得OR-MAC unit的功耗开销极低。
%(1) 由于DS-CIM2的比DS-CIM1的OR-MAC granularity更大，拥有更简单的加法逻辑和更低的累加位宽，因此DS-CIM2比DS-CIM1在MAC电路上消耗的能量更低
%（2）由于DS-CIM将2’s complement data shift到128，使得stochastic bitstream的sparsity降低，因此我们测试了DS-CIM针对有符号数的额外能量开销。测试表明DS-CIM1对符号计算比较敏感，而OR-MAC granularity更大的DS-CIM2则表现出了resilience to signed operation。
%(3) 由于共享PRNG，DS-CIM的功耗中PRNG的占比非常小。而由于大规模使用了SNG进行比较，因此SNG的动态功耗比较高。此外，大规模的accumulator的功耗开销也不可忽略。

% \begin{figure}[tp]
% \centerline{\includegraphics[width=0.8\columnwidth]{figure/fig9.png}}
% %\vspace{-2mm}
% \caption{\small{Area breakdown of DS-CIM1 and DS-CIM2, where DS-CIM2 is equipped with the latch-cached accumulator.}}
% \vspace{-3mm}
% \label{fig9}
% \end{figure}

%\textcolor{red}{The figure 8 and 9 should change the color, and the lines under pictures should be addressed.}

%在figure9中，我们展示了area breakdown of DS-CIM1 and DS-CIM2. 我们发现DS-CIM1有相当一部分的面积是adder，这导致其较高的动态功耗。这也证实了我们增大OR-MAC的granularity，减少了adder和accumulator逻辑的开销，从而使得DS-CIM2更高效。
%另外，通过增大CMR到64，我们分摊了SNG的开销，使得SNG的占比处于一个合理的区间。

%---------------------Table II------------------------------------
\begin{table*}[!ht]
    \centering
    \caption{DS-CIM Performance on FP8 Quantized LLaMA-7B Model}
    \vspace{-2mm}
    \begin{tabular}{lcccccccc}
    \hline
    \textbf{LLaMA-7B Model} & \textbf{BoolQ} & \textbf{PIQA} & \textbf{HellaSwag} & \textbf{WinoGrande} & \textbf{ARC-e} & \textbf{ARC-c} & \textbf{Avg.} & \textbf{Accuracy Degradation} \\
    \hline
    Full Precision & 75.1\% & 78.7\% & 56.9\% & 69.9\% & 75.3\% & 41.9\% & 66.3\% & 0\\
    FP8 Quantized & 75.6\% & 78.2\% & 56.6\% & 70.2\% & 74.6\% & 40.7\% & 66.0\% & 0.3\% from full precision\\
    DS-CIM1 & 71.7\% & 74.7\% & 52.1\% & 67.4\% & 70.3\% & 39.4\% & 62.6\% & 3.4\% from FP8 \\
    DS-CIM2 & 67.1\% & 70.7\% & 47.5\% & 65.4\% & 66.3\% & 35.6\% & 58.8\% & 7.2\% from FP8\\
    \hline
    \end{tabular}
    \label{Table II}
\end{table*}
%---------------------Table II------------------------------------

%------------------------Table III---------------------------------
\begin{table}[!tp]
    \centering
    %\vspace{-2mm}
    \caption{Comparison with SOTA Approximate Works}
    %\vspace{-2mm}
    \label{Table III}
    \resizebox{1.02\columnwidth}{!}{%
    \begin{threeparttable}
    \begin{tabular}{|l|c|c|c|c|c|c|}
    \hline
               & \makecell{ISSCC'23 \\ \cite{he20237}} & \makecell{DIMCA1 \\ \cite{lin2023dimca}} & \makecell{ISSCC'24 \\ \cite{yoshioka202434}} & \makecell{VLSI'24 \\ \cite{yang2024278}} & DS-CIM1 & DS-CIM2 \\ 
    \hline
    Technology & 28nm & 28nm & 65nm & 12nm & \multicolumn{2}{c|}{40nm} \\ 
    \hline
    Macro Type & Digital & Digital & Analog & \makecell{Digital \\ Stochastic} & \multicolumn{2}{c|}{\makecell{Digital Stochastic}} \\ 
    \hline
    \multirow{2}{*}{Precision} & \makecell{A: 2$\sim$8b \\ W: 2$\sim$8b} & \makecell{A: 1$\sim$4b \\ W: 1b} & \makecell{A: 4$\sim$8b \\ W: 4$\sim$8b} & \makecell{A: 2/6b \\ W: 6b} & \multicolumn{2}{c|}{\makecell{A: 8b \\ W: 8b}} \\ 
    \hline
    Array Size & 16Kb & 16Kb & 80Kb & 15Kb & \multicolumn{2}{c|}{32Kb} \\ 
    \hline
    Area (mm²) & 0.028 & 0.049 & 0.48 & 0.5 & 0.78 & 0.72 \\ 
    \hline
    Voltage (V) & 0.54$\sim$0.9 & 0.45$\sim$1.1 & 0.6$\sim$1.1 & 0.64 & \multicolumn{2}{c|}{0.7$\sim$1.2} \\ 
    \hline
    Input & \makecell{18\% \\ toggle} & \makecell{25\% \\ toggle} & N/A & N/A & \multicolumn{2}{c|}{87.5\% sparsity} \\ 
    \hline
    TOPS/mm²\textsuperscript{(1)} & 131.7 & 192.2 & 33.0 & 22.0 & \makecell{117.1\textsuperscript{(2)}\\$\sim$\textbf{468.4}\textsuperscript{(3)}} & \makecell{90.9\textsuperscript{(2)}\\$\sim$\textbf{363.7}\textsuperscript{(3)}} \\ 
    \hline
    TOPS/W\textsuperscript{(1)} & 4569.6 & 694.4 & 6652.8 & 928.8 & \makecell{669.7\textsuperscript{(2)}\\$\sim$\textbf{2677.2}\textsuperscript{(3)}} & \makecell{891.5\textsuperscript{(2)}\\$\sim$\textbf{3566.1}\textsuperscript{(3)}} \\ 
    \hline
    Retraining & N/A & Yes & N/A & N/A & \multicolumn{2}{c|}{Not Needed} \\ 
    \hline
    \makecell[l]{ResNet18/20\\@CIFAR-10\tnote{(4)}} & 91.55\% & 90.41\% & 91.70\% & N/A & \makecell{90\%\textsuperscript{(3)}\\$\sim$\textbf{94.45}\%\textsuperscript{(2)}} & \makecell{89.46\%\textsuperscript{(3)}\\$\sim$\textbf{94.31}\%\textsuperscript{(2)}} \\ 
    \hline
    \makecell[l]{ResNet50\\@ImageNet\textsuperscript{(4)}} & 74.5\% & N/A & N/A & N/A & \makecell{79.45\%\textsuperscript{(3)}\\$\sim$\textbf{80.67}\%\textsuperscript{(2)}} & \makecell{79.34\%\textsuperscript{(3)}\\$\sim$ \textbf{80.65}\%\textsuperscript{(2)}} \\ 
    \hline
    \makecell[l]{Transformer \\ Models} & N/A & N/A & \makecell{ViT-Small \\ @CIFAR10: \\ 95.8\%} & N/A & \makecell{LLaMA-7B \\@WG: \\67.4\%\textsuperscript{(2)}} & \makecell{LLaMA-7B \\@WG: \\65.4\%\textsuperscript{(2)}} \\ 
    \hline
    \end{tabular}
    \begin{tablenotes}
    \footnotesize
    \item[(1)] Scaled to 40nm, 1b input and 1b weight. 
    \item[(2)] At bitstream length=256 for best accuracy. 
    \item[(3)] At bitstream length=64 for best efficiency. 
    \item[(4)] Baseline accuracies are incomplete, and the model versions may also be different.
    \end{tablenotes}
    \end{threeparttable}
    }
    \vspace{-3mm}
\end{table}
%------------------------------Table III---------------------------

\subsection{Accuracy Evaluations}

Table \ref{Table I} shows that the RMSE is influenced by OR-MAC granularity and bitstream length, revealing a trade-off between accuracy and efficiency. The conclusions align with our initial intuitions. \textbf{First}, increasing bitstream could enhance accuracy; however, this improvement comes at the cost of degraded efficiency due to additional computing cycles. \textbf{Second}, DS-CIM1 demonstrates an accuracy advantage, while DS-CIM2 exhibits superior efficiency. \textbf{Notably}, with a bitstream length=256, DS-CIM1 achieves an accuracy of 94.45\%, with only a 0.09\% drop compared to the accurate model, while DS-CIM2 attains an accuracy of 94.31\%, reflecting a drop of 0.23\%. We also showcase DS-CIM capability on larger \texttt{ResNet50\_Weights.IMAGENET1K\_V2} on \texttt{ImageNet}, which is an improved \texttt{ResNet50} model by using a new training recipe, showing higher accuracy and robustness than previous versions. Particularly, with a bitstream length=64, DS-CIM1 experienced only a 1.37\% drop in accuracy, while DS-CIM2 had just a 1.48\% drop. This not only indicates that DS-CIM performs well on large DNNs but also highlights how continuously evolving algorithms are becoming increasingly favorable for approximate computing.

% %---------------------Figure 10------------------------------------
% \begin{figure}[tp]
% \centerline{\includegraphics[width=\columnwidth]{figure/fig10.png}}
% \vspace{-2mm}
% \caption{\small{Tradeoff analysis between accuracy, efficiency, OR-MAC granularity and bitstream length.}}
% \vspace{-5mm}
% \label{fig10}
% \end{figure}
% %---------------------Figure 10------------------------------------

% %----------fig11----------
% \begin{figure}[t]
%     \centering
%     \begin{subfigure}[t]{0.45\columnwidth} % 第一幅子图
%         \centering
%         \includegraphics[width=\columnwidth]{figure/fig11a.png} % 替换为实际图片路径
%         \vspace{-6mm}
%         \caption{DS-CIM1 layout}
%     \end{subfigure}
%     \vspace{-2mm}
%     \begin{subfigure}[t]{0.45\columnwidth} % 第二幅子图
%         \centering
%         \includegraphics[width=\columnwidth]{figure/fig11b.png} % 替换为实际图片路径
%         \vspace{-6mm}
%         \caption{DS-CIM2 layout}
%     \end{subfigure}
%     \caption{(a) The detailed layout of DS-CIM1. (b) The detailed layout of DS-CIM2.}
%     \vspace{-2mm}
%     \label{fig11}
% \end{figure}
% %----------fig11----------

%In Figure \ref{fig10}, we further explore the relationship between ResNet18 inference accuracy, hardware efficiency, OR-MAC granularity, and bitstream length.

In Table \ref{Table II}, we present the accuracy of the FP8 \texttt{LLaMA-7B} model across different datasets, along with using DS-CIM for inference. We aligns FP8 to INT8 \cite{tu2022redcim}, and applied the error pattern from DS-CIM with a bitstream length=256 to MAC outputs in the forward propagation of \texttt{LLaMA-7B}. Therefore, the error sources for DS-CIM here include two aspects: one from the process of aligning FP8 to INT8, and the other from the stochastic errors of DS-CIM. Table \ref{Table II} shows that DS-CIM maintains good accuracy in complex transformer models, where DS-CIM1 showing only a 3.4\% average accuracy drop. This demonstrates that the application scenarios for DS-CIM have been further expanded.

The analysis shows that eliminating 1s saturation errors in OR accumulation, combined with mutual cancelation of errors across different rows, leads to a uniform and stable error distribution for DS-CIM. This enables a high accuracy comparable with digital adders, making DS-CIM suitable for various application scenarios.
%结论与直觉一致，首先，增加bitstream length显著增加了accuracy，然而使用了更多的计算周期而使得efficiency降低。其次，不同的granularity之间有细微差异，DS-CIM1展现了精度方面的优势而DS-CIM2展现了efficiency方面的优势。值得一提的是，在bitstream为256时，DS-CIM1达到了94.45%的准确率，仅比准确模型下降了0.09%，DS-CIM2也达到了94.31%的准确率。分析原因可知，由于消除了OR累加误差，以及不同行之间误差的相互抵消，DS-CIM具有均匀稳定的误差分布，使得DS-CIM具有媲美accurate circuit的能力。

\subsection{Comparison with State-of-the-Art Works}

%Figure \ref{fig11} shows the implemented DS-CIM1 and DS-CIM2 macro layout overviews.
Table \ref{Table III} compares the proposed DS-CIM1 and DS-CIM2 with state-of-the-art approximate CIM designs. DS-CIM1 achieves a peak accuracy of 94.45\% with a bitstream length of 256 and a comparable energy efficiency of 669.7TOPS/W. Meanwhile, DS-CIM2 demonstrates greater power efficiency, with an energy efficiency of 3566.1 TOPS/W and an area efficiency of 363.7 TOPS/mm$^2$ at a bitstream length of 64, while maintaining an accuracy of 89.46\%. DS-CIM also supports larger models, such as \texttt{ResNet50} on \texttt{ImageNet} and the transformer model \texttt{LLaMA-7B}. Its flexible bitstream scheduling allows DS-CIM to adjust precision and efficiency for different applications.

% \begin{table}[t]
% \caption{\small{Comparison with SOTA approximate CIM macros}}
% \centerline{\includegraphics[width=\columnwidth]{figure/Table.png}}
% \vspace{-3mm}
% \label{Table}
% \end{table}

\section{Conclusion}

In conclusion, the DS-CIM architecture effectively merges high accuracy and efficiency in stochastic DCIM, overcoming challenges related to bitstream lengths and traditional DCIM limitations. By executing signed MAC operations on a unipolar OR-MAC column and replicating low-cost OR-MAC circuits, we achieve a 32$\times$ increase in compute density with only a 1$\times$ extra overhead in area. Our stochastic process analysis and data remapping techniques effectively address the 1s saturation issue in OR accumulation, enhancing accuracy and resilience to variations in sparsity. DS-CIM1 achieves an impressive ResNet18 inference accuracy of 94.45\% with a minimal RMSE of 0.74\% on the CIFAR-10 dataset, while DS-CIM2 delivers exceptional energy efficiency of 3566.1 TOPS/W, maintaining a competitive RMSE of 3.81\%. Experiments with INT8 \texttt{ResNet50} on \texttt{ImageNet} and FP8 \texttt{LLaMA-7B} also comprehensively demonstrate DS-CIM's capability with large models. This work demonstrates the potential of DS-CIM in advancing energy-efficient, high-performance, and precision-adjustable approximate computing systems.

\clearpage

\bibliographystyle{unsrt}
\bibliography{reference}

\end{document}